\documentstyle[prl,aps,preprint,tighten,floats,aps,epsf,psfig]{revtex}

\def\abstract{\if@twocolumn
\section*{Abstract}
\else \normalsize
\begin{center}
{\bf Abstract\vspace{0pt}}
\end{center}
\setlength{\baselineskip}{4ex}
\quotation
\fi}
\def\endabstract{\if@twocolumn\else\endquotation\fi}

\begin{document}
\draft
\preprint{UPR-0802-T, hep-ph/9807321}
\date{\today}
\setcounter{footnote}{1}
\title{Effects of Heavy States on the Effective $N=1$ Supersymmetric Action}
\author{Mirjam Cveti\v c, Lisa Everett, and Jing Wang}
\address{
          Department of Physics and Astronomy\\
          University of Pennsylvania\\
          Philadelphia, PA  19104\\
}

\begin{titlepage}
\maketitle
\def\thepage {}        

\begin{abstract}
Using the power of superspace formalism, we investigate the decoupling effects of
heavy states in $N=1$ supersymmetric field theory.  We find that ``mixed"
couplings in the superpotential between the heavy and light fields contribute to 
the effective superpotential at the leading order, and also contribute to the
effective K\"{a}hler potential (in the next to leading order). 
Mixed couplings in the K\"{a}hler potential always contribute to the effective
K\"{a}hler potential at the leading order.
Several examples are presented which illustrate the effects explicitly.

\end{abstract}
\end{titlepage}

\section{Introduction}

In this paper, we examine the decoupling of heavy states in {\it $N=1$
supersymmetric field theories}.  While we are motivated by
the analysis of a class of quasi-realistic string models after vacuum
restabilization~\footnote{In a class of quasi-realistic string models with an
anomalous $U(1)$, the standard anomaly cancellation mechanism generates a
Fayet-Iliopoulos contribution to the $D$- term of the anomalous $U(1)$ at
genus-one. The FI term triggers certain scalar fields to acquire vacuum
expectation values (VEV's) of ${\cal O}(M_{String}\sim 5 \times 10^{17} \, GeV)$
along $D$- and $F$- flat directions, leading to a ``restabilized" supersymmetric
string vacuum. Effective mass terms are
generated via the superpotential coupling of the fields in the model to the fields
with string-scale VEV's, and hence a number of states acquire string-scale masses
and decouple from the theory. See e.g. \cite{cceel2,cceel3} and references
therein.}, the analysis has general applications. We
specifically concentrate on the effects of gauge neutral fields.  We also do not
consider effects due to (soft) supersymmetry breaking.  

In accordance with the  well-known Applequist-Carazzone decoupling
theorem for non-supersymmetric theories \cite{appcara}, 
the tree-level exchange of heavy
fields leads to nonrenormalizable terms in the effective potential of
the light fields.  The supersymmetric
generalization as studied in \cite{katehou} reveals that the leading contribution
is to the superpotential. 
However, we also find that in general, the
decoupling of the heavy fields leads to nonrenormalizable modifications of the
K\"{a}hler potential of the light fields of the theory (as was also
pointed out in \cite{dine}).  
The effects of decoupling then lead to nonrenormalizable
interactions which are competitive at each order
with other nonrenormalizable terms present in string models (those
include nonrenormalizable terms at a given order calculated directly in
string theory and those generated from
higher-order terms which involve the fields with the large VEV's), leading to
a tower of nonrenormalizable terms to be classified in the model.

In our analysis of decoupling effects, we utilize the power of supersymmetry by
employing superspace formalism.  For the sake of simplicity, we consider the case
of gauge singlet fields. We discuss the conventional method of integrating out the
massive modes in the superspace functional integral and use the corresponding
supergraphs to illustrate the results. However, we focus on an alternative method
to determine the decoupling effects, which is to solve the equation of motion for
the heavy superfields, and then determine the effective supersymmetric action for
the light superfields of the theory. In some cases, it is possible to obtain a
complete solution to the equations of motion for the heavy fields, and thus in
principle obtain the contributions to the effective action at all orders in the
nonrenormalizable terms. 

The paper is structured as follows.  
In Section II, we develop the formalism for both methods (supergraphs and
minimization of the action) of examining the decoupling of the heavy
fields. In Section III, we investigate the effects of different
types of superpotential terms (motivated from string models) involving the heavy
fields, first taking the simplest case of a superpotential with a mass term for
the heavy fields and a term {\it linear} in the heavy field coupled to an
arbitrary function of the light fields.  
We then consider additional superpotential terms, such as a term bilinear in the
heavy fields coupled to an arbitrary function of the light fields, as well as
a trilinear self-interaction term for the heavy fields (we do not consider
higher-order interactions, as they are immediately non-renormalizable and thus
of higher order).  
We also consider the effects of non-minimal K\"{a}hler potential terms which mix
the heavy and light states.
Finally, in Section IV we present the summary and conclusions.

\section{Formalism}

The supersymmetric Lagrangian of gauge singlet chiral superfields is determined by
two functions of the
chiral superfields $\{\varphi_i\}$: (1) the K\"{a}hler
potential $K(\varphi_i,\varphi_{i}^{\dagger})$, and
(2) the superpotential $W(\varphi_i)$, where $K$ is a real function and $W$ a
holomorphic function of the chiral multiplets of the theory.
The supersymmetric action
is given by
\begin{equation}
S=\int d^{4}x d^{2}\theta d^{2}\bar{\theta} K(\varphi_{i},
\varphi_{i}^{\dagger}) + \left \{ \int d^{4}x d^{2}\theta W(\varphi_{i}) +
h.c.
\right \},
\end{equation}
in which we use the notation and conventions of \cite{wess}. 

We assume that a subset of the fields $\vec{\Phi}=\{\Phi_{l}\}$ acquire
heavy masses ${\cal O}(M)$, and
all other fields (denoted by $\vec{\varphi}=\{\varphi_{i}\}$) are light or
massless.
The K\"{a}hler potential of the theory can be written as
\begin{equation}
\label{kahler}
K=K_{min} + K',
\end{equation}
in which $K_{min}$ is the minimal canonical K\"{a}hler potential for the
theory:
\begin{equation}
\label{kahlermin}
K_{min}=\vec{\varphi}^{\dagger}\vec{\varphi}+\vec{\Phi}^{\dagger}\vec{\Phi}.
\end{equation} 
$K'$ includes possible non-minimal terms which can mix the heavy and light
fields; such terms are treated as interaction terms.

We parameterize the superpotential in powers of the heavy fields $\Phi_l$:
\begin{equation}
\label{superpot}
W=\sum_{\{l_m\}}\sum_{n=0}^{\infty}\frac{1}{n!}\Phi_{l_1}\ldots\Phi_{l_n} 
W^{\{l_1\ldots\l_n\}}_{n}(\varphi_{i});    
\end{equation}
in which the $W_{n}(\varphi_{i})$ are holomorphic functions of the light fields.  In
particular, $W_2$ includes mass terms for $\{\Phi_l\}$.  

Let us rewrite the action in term of its heavy and light components:
\begin{equation}
S(\vec{\varphi},\vec{\Phi})=S_{light}(\vec{\varphi}) 
+S_{0}(\vec{\Phi})+S_{int}(\vec{\varphi},
\vec{\Phi}). 
\end{equation}
In this expression, $S_{light}(\vec{\varphi})$ includes the free action
and the self-interactions of the light fields, such that
\begin{equation}
S_{light}(\vec{\varphi})=\int d^{4}x d^{2}\theta d^{2}\bar{\theta}
{\vec{\varphi}\vec{\varphi}^{\dagger}} +
\left \{
\int d^{4}x d^{2}\theta
W_{0}(\vec{\varphi})+h.c.  \right \}.
\end{equation} 
$S_{0}(\vec{\Phi})$ is the free action of the heavy fields
$\vec{\Phi}=\{\Phi_l\}$, and
$S_{int}(\vec{\varphi}, \vec{\Phi})$ includes the interactions between
light fields and $\vec{\Phi}$, i.e.:
\begin{equation}
\label{s0def}
S_{0}(\vec{\Phi})=\int d^{4}x d^{2}\theta d^{2}\bar{\theta} \vec{\Phi}^{\dagger}
\vec{\Phi} +\left \{\int d^{4}x d^{2}\theta
\vec{\Phi}^{T}{\bf M} \vec{\Phi}+h.c.  \right \},
\end{equation}
\begin{eqnarray}
S_{int}(\vec{\varphi}, \vec{\Phi})&=&\left \{\int d^{4}x d^{2}\theta
\sum_{\{l_m\}}
\sum_{n=1}^{\infty}\frac{1}{n!}\Phi_{l_1}\ldots\Phi_{l_n}
W^{\{l_1\ldots\l_n\}}_{n}(\varphi_{i})-\vec{\Phi}^{T}{\bf M} \vec{\Phi}
+ h.c. \right \}\nonumber\\
&+&\int d^{4}x d^{2}\theta d^{2}\bar{\theta}K',
\end{eqnarray} 
in which ${\bf M}$ is the mass matrix for $\vec{\Phi}$, with eigenvalues of ${\cal
O}(M)$. 
At energy scales lower than $M$, the heavy fields decouple from the theory, 
leading to nonrenormalizable contributions to the effective action of the light 
fields which are suppressed by inverse powers of $M$.  The conventional method to 
obtain the effective supersymmetric action is to integrate out the heavy fields 
from the theory using superspace functional integral techniques, as discussed in 
\cite{katehou}.  We briefly describe this method in part (A). However, we  
concentrate on an alternative method in this paper, by solving the full set of  
equations of motion for $\{\Phi_l\}$. We introduce this formalism in part (B).

\subsection{Superspace Functional Integral Formalism}

In general, if a supersymmetric theory contains light or massless fields 
$\{\varphi\}$ and heavy fields $\{\Phi_l\}$, the effective action 
$S_{eff}(\vec{\varphi})$ can be derived from the full action
$S(\vec{\varphi},\vec{\Phi})$ by functionally integrating over the heavy
fields: 
\begin{eqnarray}
exp(iS_{eff}(\vec{\varphi}))&=&\int [d\Phi_l] exp(iS(\vec{\varphi},
\vec{\Phi}))\nonumber\\ 
&=&exp(iS_{light}(\vec{\varphi})) \int [d\Phi_l]
exp[iS_{0}(\vec{\Phi})+iS_{int}(\vec{\varphi},\vec{\Phi})]. 
\end{eqnarray}
This expression is then rewritten as follows: 
\begin{equation}
exp(iS_{eff}(\vec{\varphi}))=exp(iS_{light}(\vec{\varphi}))exp(iS_{int}(
\frac{\delta}{\delta j_l}, \frac{\delta}{\delta j_m^\dagger})) \int
[d\Phi_l]
exp(iS_{0}(\vec{\Phi})) |_{j_l=j_m^{\dagger}=0}, 
\end{equation}
where $\vec{j}=\{j_l(z)\}$ are introduced as chiral sources for
$\vec{\Phi}$, and 
$\vec{\Phi}$ and $\vec{\Phi}^{\dagger}$ are replaced in $S_{int}$ by
$\{\frac{\delta}{\delta
j_l}\}$ and $\{\frac{\delta}{\delta j_m^\dagger}\}$, respectively. 
In the above expressions, $ \int [d\Phi_l] exp(iS_{0}(\vec{\Phi}))$ is
the free generating function
$Z_{0}[\vec{j}, \vec{j}^{\dagger}]$ for $\vec{\Phi}$ in superspace:
\begin{equation}
Z_{0}[\vec{j}, \vec{j}^{\dagger}]=exp \left( -\frac{1}{2}i\int
d^{4}xd^{4}\theta
d^{4}x^{'}d^{4}\theta^{'} [\vec{j}(z), \vec{j}^{\dagger}(z)]
\Delta_{GRS}(z, z^{'}) \left[ 
\begin{array}{c}
\vec{j}(z^{'}) \\
\vec{j}^{\dagger}(z^{'})
\end{array} \right] \right),
\end{equation}
where $z=(x, \theta, \bar{\theta})$ are the usual superspace coordinates,
$d^{4}\theta=d^{2}\theta d^{2}\bar{\theta}$, and $\Delta_{GRS}(z, z^{'})$ is
the superspace propagator given by:
\begin{equation}
\Delta_{GRS}(z, z^{'})= (\Box{\bf \rm 1}-{\bf M}^{2})^{-1}\left[ 
\begin{array}{cc}
\frac{{\bf M}D^{2}}{4\Box} & {\bf \rm 1} \\
{\bf \rm 1} & \frac{{\bf M}\bar{D}^{2}}{4\Box} 
\end{array} \right].
\end{equation}
Treating $S_{int}$ as a perturbation, the effective action can achieved via
the expansion:
\begin{equation}
exp(iS_{eff}(\vec{\varphi}))=exp(iS_{light}(\vec{\varphi}))
[1+iS_{int}(\frac{\delta}{\delta j_l}, \frac{\delta}{\delta 
j_m^\dagger}) + 
 \ldots]Z_{0}[\vec{j}, \vec{j}^{\dagger}]
|_{j_l=j_m^{\dagger}=0}. 
\end{equation}

The contributions to the effective action may arise from all orders in the expansion.  The  
corrections to the superpotential take the form of the type  
$\Pi_{i}\varphi_{i}^{n_{i}}/M^{N-3}$ (with $N=\sum_{i}n_{i}$), and the corrections
to
the K\"{a}hler potential take 
the form of the type  $\Pi_{i,j}\varphi_{i}^{n_{i}}\varphi_{j}^{\dagger n_{j}}/M^{N-2}$ (with
$N=\sum_{i,j}n_{i} n_{j}$). 

An advantage of the path integral formalism is that an inspection of
the supergraphs illustrates (and in some cases gives a systematic and compact
answer to) the new contributions to the effective action.

 
\subsection{Solving the equation-of-motion of the heavy field in superspace}

An alternative approach is to solve the full
equation of motion of the heavy fields $\{\Phi_l\}$ and substitute the solution
into the full Lagrangian to obtain the effective Lagrangian
$L_{eff}(\vec{\varphi},\vec{\varphi}^{\dagger})$ of the light fields.    

Assuming the form of the K\"{a}hler 
potential (\ref{kahler}) and the form of the superpotential given in
(\ref{superpot}), 
the equation of motion for the heavy field $\Phi_l$ is  
\begin{equation}
\label{eqmot}
-\frac{\bar{D}^{2}}{4}
\Phi_l^{\dagger}-\frac{\bar{D}^{2}}{4}\left(\frac{\partial K'}{\partial 
\Phi_l}\right)+\frac{\partial}{\partial 
\Phi_l}(\sum_{\{l_k\}}\sum_{n=1}^{\infty}\frac{1}{n!}\Phi_{l_1}\ldots\Phi_{l_n}W^{\{l_1 
\ldots l_n\}}_{n}(\varphi_{i})) =0.
\end{equation}
This equation can be solved iteratively, assuming
$\Phi_l=\sum_{n=0}^{\infty}\Phi^{(n)}_{l}$. The solution of the equation
of motion
yields an expression for $\vec{\Phi}$ as a function of $\vec{\varphi}$
and $\vec{\varphi}^{\dagger}$ in the
form of a series. This expression is then put back into the full
supersymmetric action to derive the effective  
superpotential and K\"{a}hler potential for the light fields.
As $\vec{\Phi}$ 
in general is a function of both $\vec{\varphi}$ and its conjugate,
there are terms of the form $\varphi_{i}^{n_{i}} \varphi_{j}^{\dagger
n_{j}}$ from the expansion
of the superpotential, which are corrections to the effective K\"{a}hler
potential.  Similarly, 
there are terms of form $\varphi_{i}^{n_{i}}$ from the expansion of $\vec{\Phi}^{\dagger}
\vec{\Phi}$ in the K\"{a}hler potential, which become corrections to the
effective superpotential. Hence, the new terms appearing in the effective action
can be carefully grouped to derive $K_{eff}$ and $W_{eff}$. 

In general, it is hard to obtain the full solution to the equation of
motion for $\Phi_l$.  In the next sections, we investigate several examples of
differing choices of $W$ and $K'$ involving the heavy fields in which the solution
to the equation of motion can be written in a compact form.  In one case, we find
that the corrections to the effective action can be determined to all orders in
the nonrenormalizable terms through this method, and that the calculation is in
precise agreement with the results of the calculation of the relevant supergraphs.
However, we find for most cases it is necessary to work to a certain order in the
calculation of the effective superpotential and effective K\"{a}hler potential.
We calculate the first few nonleading corrections to the effective action in these
cases.

\section{Results}

\subsection{Choices of Superpotential}

We now consider several examples of the superpotential terms that involve the
interactions between the heavy and light fields.  At present, we assume the
minimal K\"{a}hler potential (\ref{kahlermin}).  Our examples
are motivated by the types of terms that are generically present in the
superpotentials of string models.

For tree-level exchange of heavy fields, the simplest case is to assume the
presence of an interaction term of the type $\Phi_lW^{\{l\}}_1$ as well as the
relevant mass terms for the heavy fields, and an arbitrary choice of the
superpotential of the light fields.  
We then consider slightly more complicated examples, keeping in mind
that many such terms are immediately nonrenormalizable.  We choose to analyze the
case with an interaction term between one light field and two heavy fields (i.e. 
a nontrivial choice of $W_2(\vec{\varphi})$ in (4)), as well as a cubic
self-interaction among the heavy fields.

$\bullet$ {\it Example 1.}\\

We take 
\begin{equation}
\label{wex1}
W=\vec{\Phi}^{T}{\bf M}\vec{\Phi}+\vec{W}_{1}^{T}(\vec{\varphi}) \vec{\Phi}
+W_{0}(\vec{\varphi}). 
\end{equation}
In this case, $S$ is 
\begin{equation}
S=\int d^4xd^{2}\theta d^{2}\bar{\theta}  
      \vec{\Phi}^{\dagger}\vec{\Phi}+
     \left\{ \int d^4xd^{2}\theta
(\vec{\Phi}^{T}{\bf M}\vec{\Phi}+\vec{\Phi}^{T}\vec{W}_{1})
+h.c. \right\}+S_0,
\end{equation}
with $S_0$ given in (\ref{s0def}).

The equation of motion for $\vec{\Phi}$ can be written from
(\ref{eqmot}): 
\begin{equation}
-\frac{\bar{D}^{2}}{4}\vec{\Phi}^{\dagger} + \vec{\Phi}^{T}{\bf M}
+\vec{W}_{1}^{T}(\vec{\varphi})=0
\end{equation}
This set of equations can be solved iteratively, assuming
\begin{equation}
\vec{\Phi}=\sum_{n=0}^{\infty}\vec{\Phi}^{(n)},
\end{equation} 
and  that $-\frac{\bar{D}^{2}}{4}\vec{\Phi} \ll \vec{\Phi}^{T}{\bf M}
+\vec{W}_{1}(\vec{\varphi})$, such that the kinetic terms for
$\vec{\Phi}$ are
small compared to the mass terms.  The zeroth order solution is obtained
by neglecting the kinetic energy terms:
\begin{eqnarray}
\vec{\Phi}^{(0)}=-{\bf M}^{-1}\vec{W}_{1}.
\label{zero1}
\end{eqnarray}
Similarly, the first order correction is given by
\begin{equation}
\vec{\Phi}^{(1)}={\bf M}^{-1}
\frac{\bar{D}^{2}}{4}\vec{\Phi}^{(0)\,*}.
\label{first1}
\end{equation}
Repeating the procedure, the solution to the equations of motion for
$\vec{\Phi}$ takes the form of a geometric series
\begin{eqnarray}
\vec{\Phi}&=&\vec{\Phi}^{(0)}+{\bf 
M}^{-1}\frac{\bar{D}^{2}}{4}\vec{\Phi}^{(0)\,*}+
{\bf M}^{-1}\frac{\bar{D}^{2}}{4}{\bf M}^{-1}\frac{D^{2}}{4} 
\vec{\Phi}^{(0)}\nonumber\\ &+& {\bf M}^{-1} 
\frac{\bar{D}^{2}}{4}{\bf M}^{-1}
\frac{D^{2}}{4}{\bf M}^{-1}\frac{\bar{D}^{2}}{4}\vec{\Phi}^{(0)\,*} 
+\ldots \; ,
\label{series}
\end{eqnarray}
which can be summed exactly.  Using the identity
$\frac{\bar{D}^{2}D^{2}}{16}\Phi=\Box \Phi$ (for chiral superfields, which
satisfy $\bar{D}\Phi =0$), the solution can be written in the compact
form
\begin{eqnarray}
\vec{\Phi}=-({\bf M}^{2}{\bf \rm 1}-{\Box})^{-1}\left [{\bf
M}\vec{W_{1}}+\frac{\bar{D}^{2}}{4} \vec{W_{1}}^{*}\right ].
\label{soln1}
\end{eqnarray}

The solution is then substituted into the action $S$ to
determine the effective action for the light fields. 
The contributions to the effective superpotential and K\"{a}hler potential are
extracted from each term, using the identity that under an x-integration 
$d^{2}\bar{\theta}=-\frac{\bar{D}^2}{4}$ (and similarly 
$d^{2}\theta=-\frac{D^2}{4}$).

Summing up all of the contributions, the effective action 
\begin{equation}
S_{eff}=\int d^4 x d^2 \theta d^2\bar{\theta} K_{eff}+\left \{ \int d^4 x
d^2\theta W_{eff} +h.c.\right \}
\end{equation}
can be 
expressed in terms of the exact effective
K\"{a}hler potential and superpotential written in a closed form: 
\begin{eqnarray}
\label{effwgen}
W_{eff}&=&W_{0}(\vec{\varphi})-\vec{W}_{1}^{T}({\bf M}^{2}-\Box ){\bf
M} \vec{W}_1, \\
\label{effkgen}
K_{eff}&=&K_{light}+\vec{W}_{1}^{\dagger}({\bf M}^{2}-\Box )\vec{W}_{1} \ .
\end{eqnarray}
This result can be obtained using the functional integral formalism as well; the  
 relevant supergraphs are shown in Figure 1.  In this case, there is 
a $\Phi_i \Phi_j$ propagator, as well as a 
$\Phi_i \Phi_j^{ *}$ propagator.
The supergraph with the $\Phi_i \Phi_j$
propagator gives the result for the effective superpotential $W_{eff}$, while the supergraph
with the $\Phi_i \Phi_j^{*}$ propagator yields the above
expression for the effective K\"{a}hler potential $K_{eff}$.

The effective scalar potential can be derived from $W_{eff}$ and $K_{eff}$ via:
\begin{equation}
\label{scalarpot}
V_{eff}=\left[ \frac{\partial^{2}K_{eff}}{\partial \varphi_{i}\partial \varphi_{j}}
\right]^{-1}
\frac{\partial W_{eff}}{\partial \varphi_{i}}
\left(\frac{\partial W_{eff}}{\partial \varphi_{j}}\right)^{\dagger}. 
\end{equation}
Thus, the corrections to the effective scalar potential include not 
only the corrections to the superpotential, but also the effects of the 
corrections to the K\"{a}hler potential.  This effect leads to additional
nonrenormalizable terms in the scalar potential. 
The corrections from the effect of the kinetic energy of the
heavy particle, which are the terms that include the factor $\Box/M^2$, 
are higher-order derivative interactions in the theory.

To illustrate the techniques we have developed above, we consider the following
example, taking one heavy field $\Phi$ for simplicity. We assume the following
form for the superpotential, consistent with (\ref{wex1}): 
\begin{equation}
W=\frac{M}{2}\Phi^{2}+\Phi \varphi_{1}^{2}+\varphi_{1} \varphi_{2}^{2};
\end{equation}
i.e., $W_{0}=\varphi_{1} \varphi_{2}^{2}$ and $W_{1}=\varphi_{1}^{2}$. In this case,
$\Phi_{0}=-\varphi_{1}^{2}/M$. Assuming the minimal K\"{a}hler potential for the
light fields (\ref{kahlermin}), we determine the effective superpotential and
K\"{a}hler potential from (\ref{effwgen}) and (\ref{effkgen}): 
\begin{equation}
W_{eff}=\varphi_{1} \varphi_{2}^{2}-\frac{1}{1-\frac{\Box}{M^{2}}}
\frac{\varphi_{1}^{4}}{2M}; ~~~ 
K_{eff}=\varphi_{1} \varphi_{1}^{\dagger} +\varphi_{2}
\varphi_{2}^{\dagger}+\frac{1}{1-\frac{\Box}{M^{2}}}
\frac{\varphi_{1}^{2}\varphi_{1}^{2\dagger}}{M^{2}}. 
\end{equation}

To obtain the effective scalar potential, we use (\ref{scalarpot}) and neglect all
terms involving $\Box/M^2$.  The effective scalar potential for the theory is
therefore 
\begin{equation}
V_{eff}=\frac{|\varphi_{2}^{2}-\varphi_{1}^{3}/M|^{2}}{1+ 
\frac{|\varphi_{1}|^{2}}{M^{2}}}+4|\varphi_{1}\varphi_{2}|^{2}, 
\end{equation}
in which the denominator of the first term demonstrates the effect of the
modified K\"{a}hler potential.  
In particular, this effect leads to an additional term of ${\cal O}(1/M^2)$ 
in the scalar potential ($|\phi_1|^2|\phi_2|^4/M^2$).

$\bullet$ {\it Example 2.} \\

We consider the superpotential
\begin{equation}
W=W_0(\varphi_i)+\Phi W_1(\varphi_i)+\Phi^2 W_2,
\end{equation}
in which we assume only one heavy field $\Phi$ for the sake of simplicity, and
we take
\begin{equation}
W_2=\frac{M}{2}(1+\frac{2\tilde{W}_2}{M}).
\end{equation}.
The action $S$ of the heavy fields is therefore given by
\begin{equation}
S=\int d^4xd^{2}\theta d^{2}\bar{\theta}  
      (\Phi \Phi^{\dagger}) +
     \left\{ \int d^4xd^{2}\theta(\Phi^2W_2+\Phi W_{1})+h.c. \right\}.
\label{action2}
\end{equation}

The equation of motion for $\Phi$ is 
\begin{equation}
-\frac{\bar{D}^{2}}{4}\Phi^{\dagger} + W_{1}(\varphi_{i})+2\Phi W_2(\varphi_i)=0.
\end{equation}
As in the previous example, this equation can be solved iteratively in the limit
that the kinetic energy term is small.  The zeroth order solution is
\begin{equation}
\Phi^{(0)}=-\frac{W_1}{2W_2}=-(1+\frac{2\tilde{W}_2}{M})^{-1} 
\frac{W_1}{M},
\end{equation}
and the first order correction is
\begin{equation}
\Phi^{(1)}=\frac{\bar{D}^2}{8W_2}\Phi^{(0)\,\dagger}= 
(1+\frac{2\tilde{W}_2}{M})^{-1}\frac{\bar{D}^2}{4M}\Phi^{(0)\,\dagger}.
\end{equation}
Repeating the procedure to obtain the higher order terms in the expansion, the
series can be formally summed to obtain
\begin{eqnarray}
\Phi&=&\frac{1}{1-\frac{\bar{D}^2}{8W_2}\left(\frac{D^2}{8W^{\dagger}_2}\right)}\left[
\Phi^{(0)}+(1+\frac{2\tilde{W}_2}{M})^{-1} 
\frac{\bar{D}^2}{4M}\Phi^{(0)\,\dagger}\right]\nonumber\\
&=&\frac{1}{1-(1+\frac{2\tilde{W}_2}{M})^{-1}\frac{\bar{D}^2}{4M}
(1+\frac{2\tilde{W}^{\dagger}_2}{M})^{-1}\frac{D^2}{4M}} 
\left[\Phi^{(0)}+(1+\frac{2\tilde{W}_2}{M})^{-1}
\frac{\bar{D}^2}{4M}\Phi^{(0)\,\dagger}\right].
\label{soln2}
\end{eqnarray}

The form of (\ref{soln2}) indicates that in this case, it is not tractable to
extract the exact corrections to all orders in $(1/M)$ to the effective action of
the light fields.  Therefore, we work to the first nonleading contribution to the
effective action, which is the $(1/M^2)$ correction to the effective
superpotential and the $(1/M^3)$ correction to the effective K\"{a}hler potential.
Subsituting the solution (\ref{soln2}) into (\ref{action2}) and keeping only the
terms to the appropriate order, the result is
\begin{eqnarray}
W_{eff}&=&W_0-\frac{W_1^2}{2M}+\frac{W_1^2\tilde{W}_2}{M^2}+{\cal
O}\left(1/M^3\right)\\
K_{eff}&=&K_{light}+\frac{W^{\dagger}_1W_1}{M^2}+ 
2\frac{W^{\dagger}_1W_1(\tilde{W}_2+\tilde{W}^{\dagger}_2)}{M^3}+ 
{\cal O}\left(1/M^4\right).
\end{eqnarray}
The relevant supergraphs illustrate this result, and are shown in Figure 2.

$\bullet$ {\it Example 3.}\\

We consider the cubic self-interactions of one heavy field
$\Phi$, with the superpotential
\begin{equation}
W=W_0(\varphi_i)+\frac{M}{2}\Phi^2+\Phi W_1(\varphi_i)+\frac{\lambda}{3}\Phi^3,
\end{equation}
so that the supersymmetric action involving  the heavy fields is given by
\begin{equation}
S=\int d^4xd^{2}\theta d^{2}\bar{\theta} (\Phi \Phi^{\dagger}) +
     \left\{ \int d^4xd^{2}\theta(\frac{M}{2}\Phi^2+\Phi
W_{1}+\frac{\lambda}{3}\Phi^3)+h.c. \right\}.
\label{action3}
\end{equation} 
The equation of motion for $\Phi$ is
\begin{equation}
-\frac{\bar{D}^{2}}{4}\Phi^{\dagger} + W_{1}(\varphi_{i})+M\Phi+\lambda \Phi^2=0.
\end{equation}
In this case, the zeroth order iterative solution $\Phi^{(0)}$ is obtained from
the quadratic equation
\begin{equation}
\lambda \Phi^{(0)\,2}+M\Phi^{(0)}+W_1=0,
\end{equation}
with the solution
\begin{equation}
\Phi^{(0)}=-\frac{M}{2\lambda} \pm \sqrt{\left(\frac{M}{2\lambda}\right)^2- 
\left(\frac{W_1}{\lambda}\right)}.
\end{equation}
When the solution is expanded in the limit $W_1 \ll M^2$, it is evident that the
positive root is the physical solution (that corresponds to the previous result
with $\lambda=0$). The first few terms of this solution are
\begin{equation}
\label{soln03}
\Phi^{(0)}=-\frac{W_1}{M}-\frac{\lambda W_1^2}{M^3}-\frac{2\lambda^2
W_1^3}{M^5}+\ldots \; .
\end{equation}
The iterative procedure leads to the first and second order solutions
\begin{eqnarray}
\Phi^{(1)}&=&(1+\frac{2\lambda\Phi^{(0)}}{M})^{-1}
\frac{\bar{D}^2}{4M}\Phi^{(0)\,\dagger},\\
\Phi^{(2)}&=&(1+\frac{2\lambda\Phi^{(0)}}{M})^{-1}
(\frac{\bar{D}^2}{4M}\Phi^{(1)\,\dagger}-\frac{\lambda \Phi^{(1)\,2}}{M}).
\end{eqnarray}

In this case, the solution can not be summed into a compact form.  However, the
terms in the series are determined from the recursion relation
\begin{equation}
\Phi^{(n)}=(1+\frac{2\lambda\Phi^{(0)}}{M})^{-1}
(\frac{\bar{D}^2}{4M}\Phi^{(n-1)\,\dagger}-
\frac{\lambda}{M}\sum_{i=1}^{n-1}\Phi^{(i)}\Phi^{(n-i)}).
\end{equation}

Therefore, to obtain the effective supersymmetric action of the light fields, it
is necessary to work to a given order in the nonrenormalizable terms.  The
solution (\ref{soln03}) indicates that the first nonleading correction to the
superpotential from the cubic self-interaction term is of ${\cal O}(1/M^3)$, and
hence we work to that order (and to ${\cal O}(1/M^4)$ in the effective K\"{a}hler
potential).  The results are presented below:
\begin{eqnarray}
W_{eff}&=&W_0-\frac{W_1}{2M}-\frac{\lambda W_1^3}{3 M^3}+{\cal
O}\left(\frac{1}{M^5}\right)\\
K_{eff}&=&K_{light}+\frac{W_1^{\dagger}W_1}{M^2}+\frac{\lambda
W_1^{\dagger}W_1^2}{M^4}+\frac{\lambda W_1^{\dagger \,2}W_1}{M^4}+{\cal
O}\left(\frac{1}{M^6}\right).
\end{eqnarray}
We display the corresponding supergraphs in Figure 3.

\subsection{Choices of K\"{a}hler Potential}

We now consider the effects of non-minimal K\"{a}hler potential terms
involving the heavy fields. 

$\bullet$ {\it Example 1.}\\
The simplest non-trivial example in this category
is  $K'=\Phi F(\varphi_{i}, \varphi_{j}^{\dagger })+h.c.$, in
which $ F(\varphi_{i}, \varphi_{j}^{\dagger })$ is a function of either
$\varphi_{j}^{\dagger}$ only or $\varphi_{i}\varphi_{j}^{\dagger}$ (but
not a function of $\varphi_i$ only),
such that $\Phi F(\varphi_{i},\varphi_{j}^{\dagger })$ is not 
holomorphic. These terms are non-renormalizable and hence are suppressed
by $1/M^{N-2}$, where $N$ is the total power of the term.  

For this example, we take 
\begin{equation}
K=K_{light}+\Phi^{\dagger}\Phi + (\Phi F(\varphi_{i},
\varphi_{j}^{\dagger})+h.c.),
\end{equation}
\begin{equation}
W=\Phi W_{1}(\varphi_{i})+\frac{M}{2}\Phi^{2}.
\end{equation}
The supersymmetric action involving the heavy field $\Phi$ is therefore
\begin{equation}
S=\int d^4xd^{2}\theta d^{2}\bar{\theta} \left\{ \Phi \Phi^{\dagger} + (\Phi
F(\varphi_{i}, \varphi_{j}^{\dagger})+h.c.) \right\}
+   \int d^4xd^{2}\theta\left\{ (\frac{M}{2}\Phi^2+\Phi
W_{1})+h.c. \right\}.
\label{action4}
\end{equation}  
The equation of motion for $\Phi$ is 
\begin{equation}
-\frac{\bar{D}^{2}}{4}\Phi^{\dagger}-\frac{\bar{D}^{2}}{4}F+M\Phi + W_{1}=0.
\end{equation}
In this case, we can solve the equation for $\Phi$ iteratively to all
orders. However, as $\Phi F+h.c.$  is already non-renormalizable, we
only work with the lowest order solutions which lead to  non-trivial
effects in the effective action. The zeroth order solution is 
\begin{equation}
\Phi^{(0)}=-\frac{W_{1}}{M}+\frac{\bar{D}^{2}}{4M}F;
\end{equation}
and the higher order solutions can be calculated using the iteration
equation 
\begin{equation}
\Phi^{(n)}=\frac{\bar{D}^{2}}{4M} \Phi^{(n-1)\dagger}. 
\end{equation}
Keeping the lowest orders while expanding the superpotential and the
K\"{a}hler potential, the corrections to the effective action are 
\begin{eqnarray}
S_{eff} &=& S_{0}+S_{W_{1}}
\nonumber\\ 
&+& \int d^{4}x d^{2}\theta
d^{2}\bar{\theta} \left\{ (-\frac{W_{1}}{M} F - F
\frac{\bar{D}^{2}}{4M^{2}}W_{1}^{\dagger} + \frac{1}{2}
F\frac{\bar{D}^{2}}{4M}F) +h.c. + {\cal O}(1/M^{4}) \right \}.
\end{eqnarray}
$S_{W_{1}}$  includes the contributions from the superpotential term $\Phi
W_{1}$ only, which are the corrections to the effective superpotential and
K\"{a}hler potential given in (\ref{effwgen}) and (\ref{effkgen}).
 
To investigate the type of corrections in the third term of the previous
expression for $S_{eff}$, we split $F$ into its two types of terms as
follows:
\begin{equation}
F_{1}=\frac{1}{M^{N-1}}(\varphi_{1}\varphi_{2}...\varphi_{k})(\varphi_{k+1}^{\dagger}...\varphi_{N}^{\dagger});
\end{equation} 
and
\begin{equation}
F_{2}=\frac{1}{M^{N-1}}\varphi_{1}^{\dagger}\varphi_{2}^{\dagger}...\varphi_{N}^{\dagger}.
\end{equation}
As $\varphi_{i}$ are chiral fields, effectively $F_{1} \sim \varphi
\varphi^{\dagger}$ and $F_{2} \sim \varphi^{\dagger}$. We also recall that 
$W_{1}(\varphi_{i})$ is a holomorphic function of $\varphi_{i}$. 
It is clear that for the $F_1$ terms 
the highest components of $-\frac{W_{1}}{M} F$, $- F
\frac{\bar{D}^{2}}{4M^{2}}W_{1}^{\dagger}$ and $\frac{1}{2}
F\frac{\bar{D}^{2}}{4M}F$ are the $\theta \theta \bar{\theta}
\bar{\theta}$ components; hence, they are corrections to $K_{eff}$. 

For the 
$F_{2}$ terms,  the highest components of $-\frac{W_{1}}{M} F$ are the 
$\theta \theta \bar{\theta} \bar{\theta}$ components. In addition the
highest component of $- F \frac{\bar{D}^{2}}{4M^{2}}W_{1}^{\dagger}$ is 
also the $\theta \theta \bar{\theta} \bar{\theta}$ component in this case. 
To see this more clearly, note that 
\begin{equation}
\bar{D}(\bar{D}^{2}W_{1}^{\dagger})=0,
\end{equation}
as $\bar{D}$ is an anti-commuting two-component operator. Therefore,
$\bar{D}^{2}W_{1}^{\dagger}$ is effectively a chiral field of type $\varphi$,
so that $- F \frac{\bar{D}^{2}}{4M^{2}}W_{1}^{\dagger} \sim \varphi^{\dagger}
\varphi$, which has the $\theta \theta
\bar{\theta} \bar{\theta}$ component as its highest component. Similar
reasoning leads to the same result for 
$\frac{1}{2}F_{2}\frac{\bar{D}^{2}}{4M}F_{2}$.
We conclude that the leading corrections from the non-minimal K\"{a}hler
potential are corrections to the effective K\"{a}hler potential (with no direct
corrections to the
effective superpotential): 
\begin{equation}
K_{eff}=K_{light} +\frac{W_{1}^{\dagger}W_{1}}{M^{2}}-(\frac{W_{1}}{M} F + F
\frac{\bar{D}^{2}}{4M^{2}}W_{1}^{\dagger} - \frac{1}{2}
F\frac{\bar{D}^{2}}{4M}F +h.c.) + {\cal O}(1/M^{4}).   
\end{equation}
 
$\bullet$ {\it Example 2.} \\

We consider the non-minimal K\"{a}hler potential 
terms that
take the form $\Phi^{\dagger}F(\varphi_{i}, \varphi_{j}^{\dagger})\Phi$, in which  
$F(\varphi_{i}, \varphi_{j}^{\dagger})$ is a real function of the light fields
$\varphi_{i}$, $\varphi_{j}^{\dagger}$. This term is therefore suppressed by 
$1/M^{N}$, where $N$ is the total power of the light fields in $F$.  

We take  
\begin{eqnarray}
K&=&K_{light}+\Phi^{\dagger}\Phi+\Phi^{\dagger}F(\varphi_{i}, \varphi_{j})\Phi 
\\
W&=&\Phi W_{1}+\frac{M}{2}\Phi^{2}.
\label{kahler2}
\end{eqnarray}
such that the supersymmetric action for the heavy fields is 
\begin{equation}
S=\int d^4xd^{2}\theta d^{2}\bar{\theta} \left\{ \Phi \Phi^{\dagger} + \Phi
F(\varphi_{i}, \varphi_{j}^{\dagger})\Phi^{\dagger} \right\}
+   \int d^4xd^{2}\theta\left\{ (\frac{M}{2}\Phi^2+\Phi
W_{1})+h.c. \right\}.
\label{action5}
\end{equation}
The equation of motion for $\Phi$ is 
\begin{equation}
-\frac{\bar{D}^{2}}{4}(\Phi^{\dagger})-\frac{\bar{D}^{2}}{4}(F\Phi^{\dagger})+M\Phi+W_{1}=0.
\end{equation}
As $F$ is of higher order in $1/M$, the zeroth order solution is 
\begin{equation}
\Phi^{(0)}=-\frac{W_{1}}{M}; 
\end{equation}
the higher order solutions can be calculated using the iteration equation
\begin{equation}
\Phi^{(n)}=\frac{\bar{D}^{2}}{4M}\Phi^{(n-1)\dagger}+\frac{\bar{D}^{2}}{4M}(F\Phi^{(n-1)\dagger}).
\end{equation}
To demonstrate the non-trivial effects of $F$, we retain the terms up to third  
order in $(1/M)$ in the expansion of the superpotential and the K\"{a}hler
potential. As in the previous example. the non-minimal K\"{a}hler potential term
contributes to the effective K\"{a}hler potential only (and not the effective
superpotential): 
\begin{eqnarray}
K_{eff} &=& K_{light}+\frac{W_{1}^{\dagger}W_{1}}{M^{2}}+\frac{W_{1}FW_{1}^{\dagger}}{M^{2}}
\nonumber \\
 &+& 
\frac{1}{M^{3}}[W_{1}^{\dagger}\frac{\bar{D}^{2}}{4}(FW_{1}^{\dagger})+(FW_{1}^{\dagger})\frac{\bar{D}^{2}}{4}W_{1}^{\dagger}+ 
(FW_{1}^{\dagger})\frac{\bar{D}^{2}}{8}(FW_{1}^{\dagger})
+h.c.]+ {\cal O}(1/M^{4}).
\end{eqnarray}
Note that the last term in the set of $1/M^{3}$ terms has two factors of $F$,
and thus is potentially of higher order than the other two terms. 
The effective superpotential includes contributions from the term $\int d^{4}x 
d^{2}\theta \Phi W_{1}$ as usual, which is given in (\ref{effwgen}).   

This example has relevance for the issue of decoupling in gauge theories as well.
In this case, $F$ is a function of the corresponding vector supermultiplets. 
For example, in the case of an Abelian gauge theory in which $\Phi$ has $U(1)$ charge
$q$, the gauge invariant kinetic energy term for $\Phi$ 
is of the type $\Phi ^{\dagger}e^{qgV} \Phi^{\dagger}$, where $g$ is the gauge coupling
and $V$ denotes the vector supermultiplet. In the Wess-Zumino gauge,
$F(V)=e^{qgV}-1=qgV+q^{2}g^{2}V^{2}/2$, and we can apply
the techniques we have developed previously. We find that the lowest order
correction to the effective superpotential involving $F(V)$ is $\int d^{4}
d^{2}\theta 
d^{2}\bar{\theta}(1/M^{2})W_{1}(qgV+q^{2}g^{2}V^{2}/2)W_{1}^{\dagger}$.
In contrast to $F(\varphi_{i}, \varphi_{j}^{\dagger})$, $F(V)$ does not have
additional suppressions in powers of $(1/M)$. Hence, the lowest order correction is
comparable to the lowest contribution from $\Phi W_{1}$ in the superpotential.

\section{Conclusions}

We addressed the effects of  the decoupling of heavy fields in the
$N=1$ supersymmetric action. While the study  of these effects is motivated by
vacuum restabilization  of string vacua due to an anomalous  $U(1)$, the
discussion is given in a general context of $N=1$ supersymmetric field theories
(of gauge singlet fields).

We employed an iterative procedure to  solve equations of motion for the
heavy chiral superfields ${\vec \Phi}$, which allows for a controlled
study of the corrections at each order to both the effective superpotential and
K\"{a}hler potential; in some specific cases, in particular the case
 with the  ``mixed'' couplings between the heavy and light  chiral superfields ${\vec \phi}$
 of the type $ {\vec \Phi}^T{\vec W}_1({\vec \phi})$,  the full summation is possible. This method is also  illustrated  in a complementary way by employing
 the corresponding supergraphs.

  For specific examples of mixed couplings  appearing in the
 superpotential, we demonstrated that the leading correction is indeed 
 to the superpotential \cite{katehou}. However, the next-order corrections
 are not only to the superpotential but also   to
 the  K\"{a}hler potential, and signify new  effects to the
 effective potential of the theory. In general, the corrections to the K\"{a}hler
potential are one order higher than the corrections to the effective
superpotential of the theory. 

 In examples with the   mixed couplings  arising in the K\"{a}hler
 potential we  found that these terms always contribute to the leading
 corrections in the effective K\"{a}hler potential  (and are generically of the
higher order then the leading correction due  to the mixed couplings appearing in
the superpotential). 

In the case of supersymmetric gauge field theories, corrections due to the 
coupling of the heavy fields to gauge fields naturally appear in the  K\"{a}hler
potential. We found that these effects
 again  correct the effective K\"{a}hler potential (in the leading order);
 however, further study that may explore possible corrections to the
 effective gauge function (specifying the effective gauge coupling) is underway.

\section{Acknowledgments}

 This work was supported in part by U.S. Department of Energy Grant No.
DOE-EY-76-02-3071.  We thank J. R. Espinosa for his participation in the initial
stages of the collaboration, and for many helpful discussions. 
We also thank P. Langacker, K. Dienes, V. Kaplunovsky, and M. Porrati for
useful discussions.


\bibliographystyle{prsty}
\bibliography{tdw}

\newpage
\begin{figure}
\vskip -2 truein
\centerline{
\hbox{
\epsfxsize=2.8truein
\epsfbox[70 32 545 740]{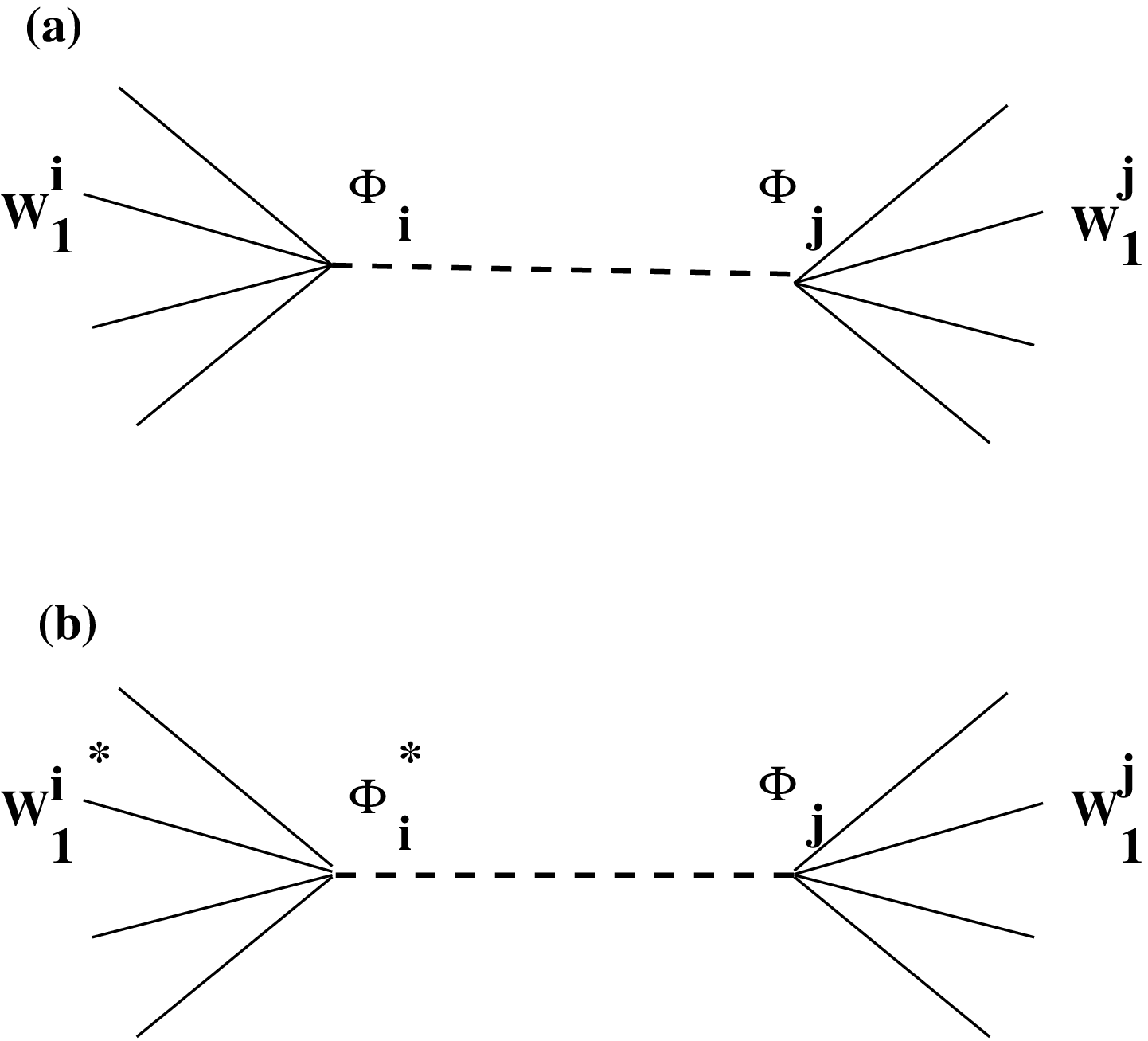}
}
}
\vskip 0.5truein
\caption{The supergraphs contributing to (a) the effective superpotential, and (b)
the effective K\"{a}hler potential for Example 1, with $W=\vec{\Phi}^T \vec{W}_1$. 
}
\vskip 0.1truein
\centerline{
\hbox{
\epsfxsize=2.8truein
\epsfbox[70 32 545 740]{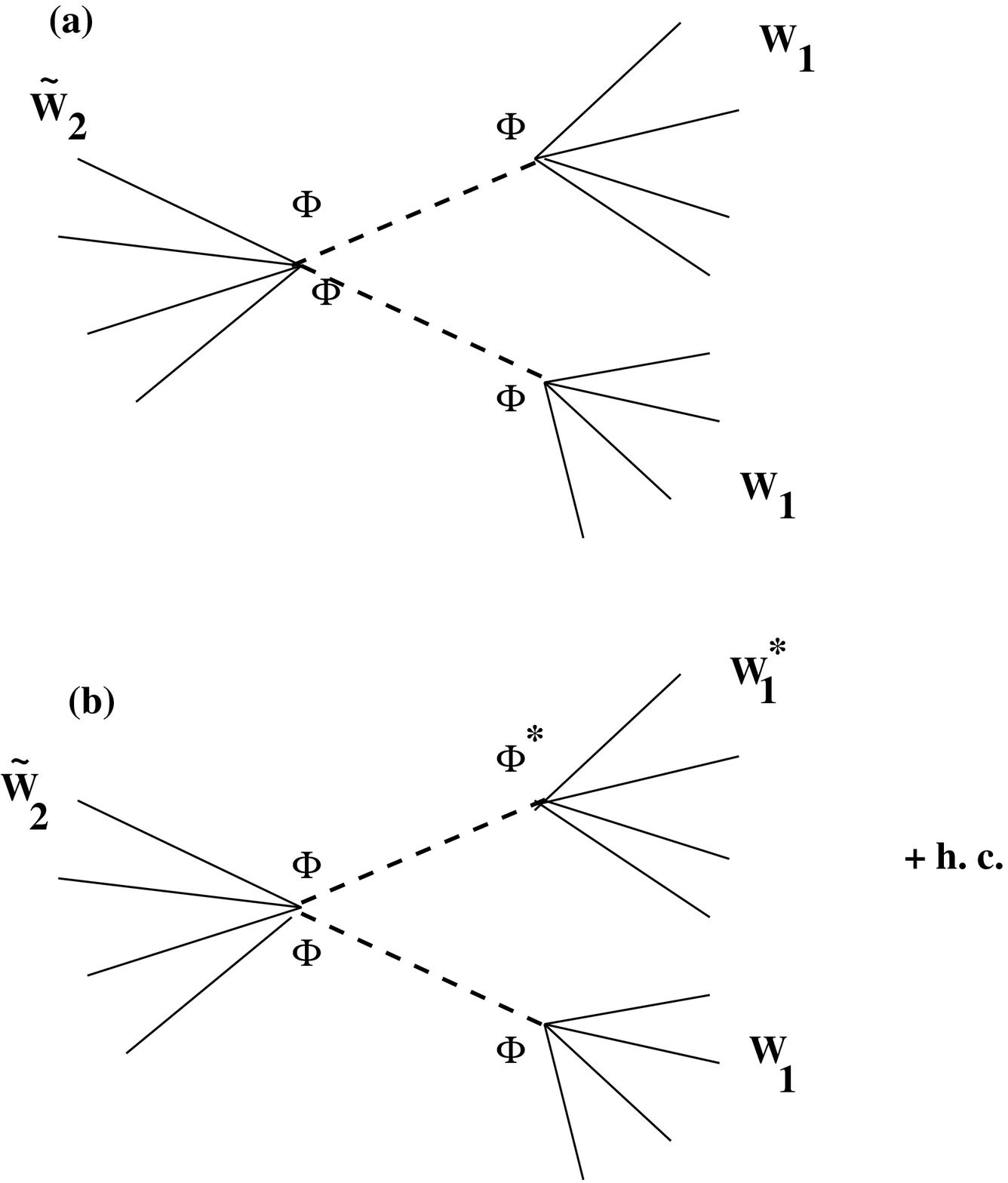}
}
}
\vskip 0.5truein
\caption{The supergraphs contributing to (a) the effective superpotential, and (b)
the effective K\"{a}hler potential for Example 2, with $W=\Phi W_1+\Phi^2
W_2$.
}
\end{figure}
\newpage
\begin{figure}
\vskip -2 truein
\centerline{
\hbox{
\epsfxsize=2.8truein
\epsfbox[70 32 545 740]{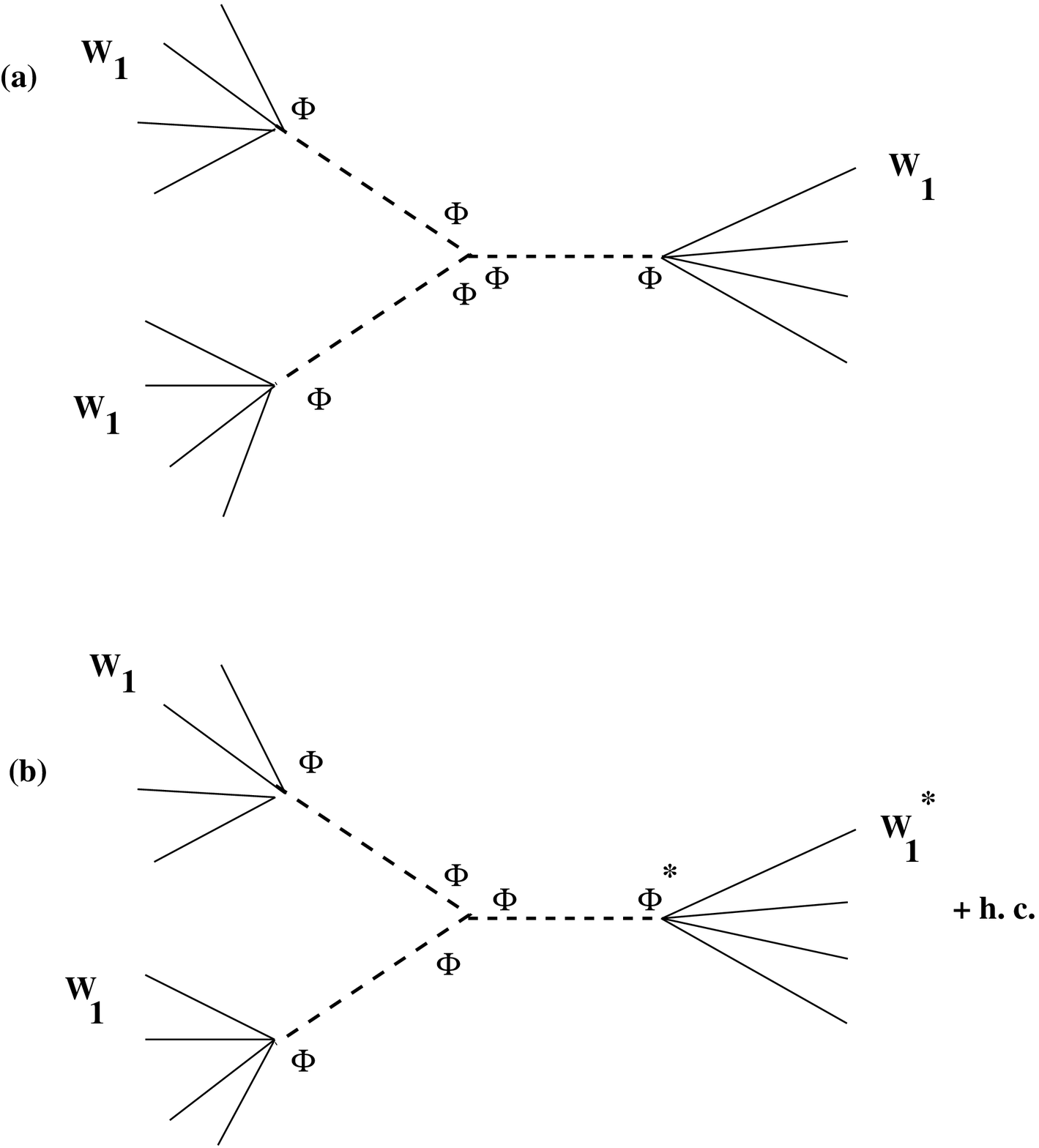}
}
}
\vskip 0.5 truein
\caption{The supergraphs contributing to (a) the effective superpotential, and
(b) the effective K\"{a}hler potential for Example 3, with $W=\Phi W_1+\Phi^3$.
}
\end{figure}

\end{document}